\documentclass[prd,superscriptaddress,amsfonts,amssymb,amsmath,showpacs,onecolumn]{revtex4-2}

\usepackage{bm}
\usepackage{amsfonts}
\usepackage{latexsym}
\usepackage{graphicx}
\usepackage{amsmath}
\usepackage{palatino}
\usepackage{mathpazo}
\usepackage{xcolor}
\usepackage{textcomp}
\usepackage{float}
\usepackage{booktabs}
\usepackage{dcolumn}
\usepackage{booktabs}
\usepackage{multirow}
\usepackage{hyperref}
\usepackage{orcidlink}
\hypersetup{colorlinks,citecolor=blue}
\usepackage[caption=false]{subfig}
\usepackage{xcolor}

\begin{document}
	\color{black}       
	\title{Viscous Ghost Dark Energy in $f(Q, L_m)$ Gravity: Observational and Thermodynamic Validity}
		
	\author{Ajay Kumar\orcidlink{0000-0002-9478-2175}}
	\email{maths.ajy@gmail.com}
	\affiliation{School of Technology Management \& Engineering,
		SVKM's NMIMS (Deemed to be University),
		Chandigarh - 160014, India.}
	
	\author{Pankaj Kumar\orcidlink{0000-0002-7156-5637}}
	\email{pankaj.11dtu@gmail.com}
	\affiliation{Department of Mathematics, Maharaja Agrasen University, Kalujhanda, Himachal Pradesh-174103, India}

\begin{abstract}
We investigate the cosmological implications of viscous ghost dark energy in the framework of modified $f(Q,\mathcal{L}_m)$ gravity by considering the linear functional form $f(Q,\mathcal{L}_m)=\alpha Q+\beta\mathcal{L}_m$. The non-minimal coupling between geometry and matter together with bulk viscous effects modifies the cosmic dynamics, leading to an analytical solution of the background Friedmann equations. Since the expansion history depends only on effective combinations of the model parameters, the observational analysis is performed in terms of the corresponding effective background parameters. We constrain the model using the latest joint compilation of Pantheon+ including SH0ES Type Ia supernovae, Cosmic Chronometers, and DESI baryon acoustic oscillation measurements. The model provides a satisfactory description of the late-time expansion history and reproduces the observed transition from decelerated to accelerated cosmic expansion. A statistical comparison with the standard $\Lambda$CDM model is performed using the minimum $\chi^2$, Akaike Information Criterion (AIC), and Bayesian Information Criterion (BIC). Furthermore, the thermodynamic viability of the model is examined through the evolution of the total entropy, and the generalized second law of thermodynamics is found to remain valid throughout the cosmic evolution.

\end{abstract}

\maketitle

\section{Introduction}

The discovery of late-time accelerated expansion of the Universe through observations of Type Ia supernovae (SNe Ia)~\cite{Riess1998,Perlmutter1999}, later confirmed by baryon acoustic oscillation (BAO) measurements~\cite{Eisenstein2005,Tegmark2004} and the cosmic microwave background (CMB) anisotropies~\cite{Spergel2007,Planck2020}, represents one of the most remarkable developments in modern cosmology. Although the standard cosmological model successfully explains a wide range of observations, the physical origin of the accelerated expansion remains unknown. Consequently, understanding the nature of dark energy (DE) and the underlying theory of gravitation continues to be one of the most challenging problems in theoretical physics.

The standard cosmological model ($\Lambda$CDM) attributes the late-time acceleration to a cosmological constant $\Lambda$ within the framework of General Relativity (GR), where the DE equation-of-state parameter is fixed at $\omega_\Lambda=-1$~\cite{Weinberg1989}. Despite its remarkable observational success, the $\Lambda$CDM model suffers from several well-known theoretical difficulties, including the cosmological constant and coincidence problems~\cite{Weinberg1989,Copeland2006,Sahni2014}. These limitations have motivated extensive investigations of dynamical DE models and modified theories of gravity, which provide alternative explanations for the observed cosmic acceleration without relying exclusively on a cosmological constant.

Among the various dynamical DE candidates, Ghost Dark Energy (GDE) has attracted considerable attention because of its origin in the low-energy effective theory of Quantum Chromodynamics (QCD)~\cite{Urban2010,Ohta2011}. In this framework, the DE density originates from the Veneziano ghost field~\cite{Veneziano1979} and is assumed to be proportional to the Hubble parameter,
\begin{equation}
	\rho_{\rm GDE}=\gamma H,
\end{equation}
where $\gamma$ is a constant related to the QCD energy scale through $\gamma\sim\Lambda_{\rm QCD}^{3}$~\cite{Ohta2011,Cai2012}. Since the DE density evolves with the cosmic expansion, GDE provides a simple dynamical alternative to the cosmological constant without introducing additional scalar degrees of freedom.

An alternative approach to explaining the accelerated expansion is to modify the gravitational sector itself. Several modified gravity theories have been proposed in this direction, including $f(R)$ gravity~\cite{nojiri2011}, $f(R,T)$ gravity~\cite{harko2011}, $f(G)$ gravity~\cite{felice2009}, $f(Q)$ gravity~\cite{jimenez2018}, and $f(Q,T)$ gravity~\cite{xu2019}. Among these, symmetric teleparallel gravity attributes gravitation to the non-metricity scalar $Q$ instead of curvature or torsion~\cite{Nester1998,Adak2006,Jarv2018}. Recently, the $f(Q,\mathcal{L}_m)$ theory has been proposed as an extension in which the gravitational action depends on both the non-metricity scalar $Q$ and the matter Lagrangian density $\mathcal{L}_m$~\cite{Hazarika2025}. The explicit coupling between geometry and matter modifies the gravitational field equations, leads to a non-conserved energy--momentum tensor, and enriches the cosmological dynamics, making the theory an attractive framework for investigating the late-time evolution of the Universe.

Besides modified gravity, dissipative processes may also influence cosmic evolution. Bulk viscosity naturally generates an effective negative pressure while preserving the homogeneity and isotropy of the Friedmann-Lemaître-Robertson-Walker spacetime. Physically, bulk viscosity arises when the cosmological fluid departs from local thermodynamic equilibrium during the expansion of the Universe~\cite{ilg1999,wilson2007}. Although the first-order Eckart formalism~\cite{eckart1940} has been widely employed in cosmological applications, it suffers from well-known causality and stability issues. These shortcomings were addressed in the second-order Israel--Stewart theory~\cite{israel1979}, while the Eckart description is recovered in
the vanishing relaxation-time limit. Owing to its mathematical simplicity and applicability to late-time cosmology, the Eckart framework continues to provide useful phenomenological descriptions of bulk viscous fluids. Numerous studies have explored different parameterizations of bulk viscosity in cosmology~\cite{fabris2006,ren2006,meng2007,singh2007a,singh2007b,avelino2008,avelino2010,avelino2013,singh2014,singh2018a,singh2018b,singh2019,singh2021,singh2023,singh2024,singh2025}, while several recent investigations have considered viscous cosmological models in $f(Q,\mathcal{L}_m)$ gravity~\cite{Myrzakulov2024a,Myrzakulov2024b,Myrzakulov2025,Rana2025}.

Despite the growing literature on viscous cosmology and modified
gravity, the specific combination of GDE, bulk viscosity, and $f(Q,\mathcal{L}_m)$ gravity considered in this work has not been explored in the existing studies. In particular, Ref.~\cite{Rana2025}
investigated viscous cosmology in $f(Q,\mathcal{L}_m)$ gravity using
Cosmic Chronometers, BAO, and GRB observations, but did not consider a GDE prescription or the thermodynamic viability of the
resulting cosmological evolution. The present work extends this framework by incorporating GDE, characterized by
$\rho_{\rm GDE}=\gamma H$, together with bulk viscosity and the non-minimal geometry--matter coupling of $f(Q,\mathcal{L}_m)$ gravity.
In addition to constraining the resulting effective background parameters using the Pantheon+SH0ES Type Ia supernova compilation,
Cosmic Chronometer measurements, and DESI BAO observations, we assess the model through a statistical comparison with $\Lambda$CDM using
$\chi^2_{\rm min}$, AIC, and BIC. Finally, we investigate the thermodynamic viability of the model through the generalized second law of thermodynamics.

The remainder of this paper is organized as follows. In Section 2, we
briefly review the field equations of $f(Q,\mathcal{L}_m)$ gravity.
Section 3 develops the viscous ghost dark energy model and derives the
corresponding cosmological solution. The observational datasets and
Bayesian parameter estimation methodology are presented in Sections 4
and 5, respectively. Section 6 discusses the cosmological evolution
and physical viability of the model, including the deceleration
parameter, effective equation of state, matter density evolution,
Hubble parameter, and thermodynamic behavior in Sections 6.1--6.5,
respectively. Finally, the main conclusions are summarized in
Section 7.

\section{$f(Q, \mathcal{L}_m)$  Gravity Theory}
Gravity is described by the curvature of spacetime in GR, where the Levi--Civita connection is metric compatible and torsion free. In the more general metric-affine formulation, however, the metric tensor $g_{\mu\nu}$ and the affine connection $\Upsilon^\alpha_{\ \mu\nu}$ are treated as independent geometrical variables, allowing the spacetime to possess both torsion and non-metricity. In this framework, the affine connection can be decomposed into three fundamental components~\cite{Hehl1995}
\begin{equation}\label{eq1}
	\Upsilon^\alpha_{\ \mu\nu} = \Gamma^\alpha_{\ \mu\nu} + K^\alpha_{\ \mu\nu} + L^\alpha_{\ \mu\nu},
\end{equation}
where $K^\alpha_{\ \mu\nu}$ is the contortion tensor associated with torsion, $L^\alpha_{\ \mu\nu}$ is the disformation tensor associated with non-metricity, and $\Gamma^\alpha_{\ \mu\nu}$ is the Levi--Civita connection, which is defined as
\begin{equation}\label{eq2}
	\Gamma^\alpha_{\ \mu\nu} = \frac{1}{2} g^{\alpha\lambda}
	\left(
	\partial_\mu g_{\lambda\nu} +
	\partial_\nu g_{\lambda\mu} -
	\partial_\lambda g_{\mu\nu}
	\right).
\end{equation}
The contortion tensor is related to the torsion tensor through
\begin{equation}\label{eq3}
	K^\alpha_{\ \mu\nu} =
	\frac{1}{2}
	\left(
	T^\alpha_{\ \mu\nu}
	+
	T_{\mu\ \nu}^{\ \alpha}
	+
	T_{\nu\ \mu}^{\ \alpha}
	\right).
\end{equation}
The disformation tensor, which encodes the effects of non-metricity, is defined as
\begin{equation}\label{eq4}
	L^\alpha_{\ \mu\nu} =
	\frac{1}{2}
	\left(
	Q^\alpha_{\ \mu\nu}
	-
	Q_{\mu\ \nu}^{\ \alpha}
	-
	Q_{\nu\ \mu}^{\ \alpha}
	\right),
\end{equation}
where the non-metricity tensor is defined as
\begin{equation}\label{eq5}
	Q_{\alpha\mu\nu} = \nabla_\alpha g_{\mu\nu}.
\end{equation}
These geometrical quantities constitute the basic building blocks of metric-affine gravity and, in particular, symmetric teleparallel gravity, where gravitation is entirely described by non-metricity rather than curvature or torsion~\cite{jimenez2018}.

Symmetric Teleparallel Equivalent of General Relativity (STEGR) is obtained by imposing the conditions
\begin{equation}
T^\alpha_{\ \mu\nu}=0,\qquad
R^\alpha_{\ \beta\mu\nu}=0,
\end{equation}
such that gravitation is completely described by the non-metricity scalar $Q$~\cite{jimenez2018}. This framework forms the basis of both $f(Q)$ and $f(Q,\mathcal{L}_m)$ gravity theories.

Two vector contractions constructed from the non-metricity tensor are given by
\begin{equation}\label{eq6}
	Q_\alpha = Q_{\alpha\mu}{}^\mu, \qquad \tilde{Q}_\alpha = Q^\mu_{\ \alpha\mu}.
\end{equation}
The corresponding superpotential is defined as
\begin{equation}\label{eq7}
	P^\alpha_{\ \mu\nu} =
	-\frac{1}{2} L^\alpha_{\ \mu\nu}
	+ \frac{1}{4} (Q^\alpha - \tilde{Q}^\alpha) g_{\mu\nu}
	- \frac{1}{4} \delta^\alpha_{(\mu} Q_{\nu)},
\end{equation}
which is analogous to the superpotential appearing in teleparallel gravity. Using these quantities, the non-metricity scalar is defined as
\begin{equation}\label{eq8}
	Q = -Q_{\lambda\mu\nu} P^{\lambda\mu\nu}.
\end{equation}
The action of $f(Q,\mathcal{L}_m)$ gravity is written as ~\cite{Hazarika2025},
\begin{equation}\label{eq9}
	S = \int d^4x \sqrt{-g}\, f(Q, \mathcal{L}_m),
\end{equation}
where  the function $f(Q, \mathcal{L}_m)$  is an arbitrary function of the non-metricity scalar $Q$ and the matter Lagrangian $\mathcal{L}_m$, introducing modifications beyond standard teleparallel and GR formulations.

In the present work we adopt a linear functional form
\begin{equation}
f(Q,\mathcal L_m)=\alpha Q+\beta\mathcal L_m,
\end{equation}
where $\alpha$ and $\beta$ are constant model parameters. This choice leads to analytically tractable cosmological field equations while retaining the non-minimal coupling between geometry and matter.

Variation of the action with respect to the metric tensor yields the modified gravitational field equations
\begin{equation}\label{eq10}
	\frac{2}{\sqrt{-g}} \nabla_\alpha \left( \sqrt{-g} f_Q P^\alpha_{\ \mu\nu} \right) + f_Q \left( P_{\mu\alpha\beta} Q^{\alpha\beta}{}_\nu - 2Q_{\alpha\beta\mu} P^{\alpha\beta}{}_\nu \right) + \frac{1}{2}fg_{\mu\nu} = \frac{1}{2} f_{\mathcal{L}_m} \left( g_{\mu\nu}\mathcal{L}_m - T_{\mu\nu} \right),
\end{equation}
where ${f_Q=\partial f(Q, \mathcal{L}_m)/\partial Q}$ and ${f_{\mathcal{L}_m}=\partial f(Q, \mathcal{L}_m)/\partial \mathcal{L}_m}$. The matter energy-momentum tensor is defined as
\begin{equation}\label{eq11}
	T_{\mu\nu} = g_{\mu\nu}\mathcal{L}_m - 2\frac{\partial \mathcal{L}_m}{\partial g^{\mu\nu}}.
\end{equation}
Unlike GR, where $\nabla_\mu T^{\mu\nu}=0$ automatically, variation with respect to the affine connection yields additional dynamical equations leading to a non-conservation law
\begin{equation}\label{eq12}
	\mathcal{D}_\mu T^\mu_{\ \nu} \neq 0.
\end{equation}
Equation (\ref{eq12}) implies that the energy--momentum tensor is generally not conserved because of the explicit coupling between geometry and matter. Consequently, an exchange of energy and momentum occurs between the gravitational and matter sectors, leading to an extra force acting on massive test particles.

It is worth emphasizing that the linear form adopted in the present work admits an analytical solution for the background expansion history. As shown in the following section, the resulting Hubble parameter depends only on a small number of effective combinations of the underlying theoretical parameters. Consequently, the observational analysis is naturally formulated in terms of these effective background parameters, rather than attempting
to constrain the underlying degenerate parameters individually.

\section{Viscous Ghost Dark Energy in $f(Q,\mathcal{L}_m)$ Gravity}
For the spatially flat Friedmann--Lema\^{i}tre--Robertson--Walker (FLRW) spacetime
\begin{equation}
	ds^2=-dt^2+a^2(t)(dx^2+dy^2+dz^2),
\end{equation}
the non-metricity scalar reduces to
\begin{equation*}
Q=6H^2,
\end{equation*}
where $H=\frac{\dot a}{a}$ is the Hubble parameter describing the cosmic expansion rate. Throughout this work, an overdot denotes differentiation with respect to the cosmic time $t$.

The energy-momentum tensor comprises of energy density $\rho$ and pressure $\bar{p}$ of the cosmic fluid with viscosity effect is given by \cite{landau1987}
\begin{equation}\label{eq14}
	T_{\mu\nu} = (\rho + \bar{p})u_\mu u_\nu + \bar{p} g_{\mu\nu},
\end{equation}
where $\bar{p} = p - 3\zeta H$ and $u^{\mu}=(1,0,0,0)$ are the components of the four velocities. Here, $\rho=\rho_m+\rho_d$ denotes the total energy density of the cosmic fluid, while the equilibrium pressure is $p=p_m+p_d$. Since dark matter is assumed to be pressureless, $p_m=0$, the pressure is entirely provided by the GDE component. In the presence of bulk viscosity, the effective pressure becomes $\bar p=p_d-3\zeta H$, where $\zeta$ is the bulk viscosity coefficient.

Assuming a barotropic equation of state for the ghost dark energy, $p_d=\omega\rho_d$, the effective pressure takes the form
\begin{equation}\label{eq15}
	\bar p=\omega\rho_d-3\zeta H.
\end{equation}
Following Refs.~\cite{Singh2008,Meng2009}, we adopt the linear parametrization of the bulk viscosity coefficient in the form
\begin{equation}\label{eq16}
	\zeta = \zeta_0 + \zeta_1 H,
\end{equation}
where $\zeta_0$ and $\zeta_1$ are constants. This choice is both theoretically well motivated and phenomenologically justified. In relativistic fluid dynamics, bulk viscosity arises due to deviations from local thermodynamic equilibrium during the expansion of a fluid. The strength of such dissipative processes is governed by the scalar expansion $\theta=\nabla_\mu u^\mu$, which, for a homogeneous and isotropic FLRW universe, reduces to $\theta=3H$. Consequently, it is natural to model the bulk viscosity as a function of the Hubble parameter, which characterizes the cosmic expansion rate.

Even in regimes of modest expansion, intrinsic dissipative effects that could be related to microscopic processes like particle creation, vacuum decay, or internal interactions of the cosmic fluid are taken into account by the constant term $\zeta_0$. Expansion-induced viscous effects are captured by the term proportional to $H$, denoted as $\zeta_1 H$, which reflects the enhancement of irreversible processes in a rapidly expanding universe. As a result, this combined form naturally includes the often studied instances $\zeta=\zeta_0$ and $\zeta\propto H$ as limiting scenarios and constitutes the most general linear parametrization of bulk viscosity in terms of the expansion rate. Thermodynamic consistency requires the effective bulk viscosity coefficient to remain non-negative, i.e., $\zeta(H)=\zeta_0+\zeta_1H\geq0$ throughout the cosmological evolution considered. 

Substituting the FLRW metric together with the viscous fluid energy--momentum tensor into the field equations of $f(Q,\mathcal{L}_m)$ gravity yields the modified Friedmann equations governing the cosmological evolution of a bulk viscous ghost dark energy dominated Universe~\cite{Hazarika2025}:
\begin{equation}\label{eq17}
3H^2=\frac{1}{4f_{Q}}\left[f-f_{\mathcal{L}_m}(\rho+\mathcal{L}_m)\right],
\end{equation}
\begin{equation}\label{eq18}
\dot{H}+3H^2+\frac{\dot{f}_{Q}}{f_{Q}}H=\frac{1}{4f_{Q}}\left[f+f_{\mathcal{L}_m}(\bar{p}-\mathcal{L}_m)\right].
\end{equation}
 To ensure compatibility with GR in the weak-field limit and avoid instability due to higher-order corrections, we adopt a linear form of the gravitational Lagrangian,
\begin{equation}
	f(Q, \mathcal{L}_m) = \alpha Q + \beta \mathcal{L}_m,
\end{equation}
which represents the simplest extension incorporating a non-minimal coupling between geometry and matter while preserving linearity in both $Q$ and $\mathcal L_m$. Using $p_m=0$, together with $\mathcal L_m=\bar p$, the modified Friedmann equations reduce to
\begin{equation}\label{eq19}
	3H^2 + \frac{\beta}{2\alpha}\rho = 0,
\end{equation}
\begin{equation}\label{eq20}
	2\dot{H} + 3H^2 - \frac{\beta}{2\alpha}\bar{p}=0.
\end{equation}
Following the ghost dark energy scenario, the dark energy density is assumed to be proportional to the Hubble parameter, $\rho_d=\gamma H$, where $\gamma$ is a constant with dimensions of energy cubed and is expected to be of the order of $\Lambda_{\rm QCD}^3$, with $\Lambda_{\rm QCD}\sim100\,{\rm MeV}$ representing the QCD energy scale.

Substituting $\rho_d=\gamma H$, together with $p_d=\omega\rho_d$ and the viscosity parametrization $\zeta=\zeta_0+\zeta_1H$, into Eqs.~\eqref{eq19} and \eqref{eq20}, we obtain
\begin{equation}\label{eq23}
	\dot{H} + \frac{3}{4\alpha}(2\alpha+\beta \zeta_1)H^2 - \frac{\beta}{4\alpha}(\gamma\omega - 3\zeta_0)H = 0.
\end{equation}
Using the relation $z=a^{-1}-1$, Eq.\eqref{eq23} can be rewritten as
\begin{equation}\label{eq24}
	\frac{dH}{dz} - \frac{3}{4\alpha}(2\alpha+\beta \zeta_1)(1+z)^{-1}H + \frac{\beta}{4\alpha}(\gamma \omega - 3\zeta_0)(1+z)^{-1} = 0.
\end{equation}
Solving Eq.~(26), we obtain the following expression for the Hubble parameter as a function of redshift.
\begin{equation}\label{eq25}
	H(z)=H_0\left[\frac{\beta (\gamma\omega - 3\zeta_0)}{3H_0(2\alpha+\beta \zeta_1)} +
	\left( 1 - \frac{\beta (\gamma\omega - 3\zeta_0)}{3H_0(2\alpha+\beta \zeta_1)}\right)(1+z)^{\frac{3(2\alpha+\beta \zeta_1)}{4\alpha}}\right],
\end{equation}
where $H_0$ is the present value of the Hubble parameter.

For convenience, we introduce the effective parameter combinations
\begin{equation}\label{eq28}
	m=\frac{\beta(\gamma\omega-3\zeta_0)}
	{3H_0(2\alpha+\beta\zeta_1)},
	\qquad
	n=\frac{3(2\alpha+\beta\zeta_1)}
	{4\alpha},
\end{equation}
so that the Hubble parameter can be written in the compact form
\begin{equation}\label{eq29}
	H(z)=H_0\left[m+(1-m)(1+z)^n\right].
\end{equation}
It is evident from Eq.~(\ref{eq29}) that the background expansion history depends only on the effective parameter combinations $m$ and $n$, rather than on the individual model parameters $\alpha$, $\beta$, $\gamma$, $\omega$, $\zeta_0$, and $\zeta_1$. It is important to note that this reparameterization does not eliminate the physical effects of bulk viscosity or the non-minimal geometry--matter coupling. Rather, their effects on the background expansion are encoded in the effective combinations $m$ and $n$. In particular, $\zeta_0$ enters through the combination $(\gamma\omega-3\zeta_0)$, while $\zeta_1$ contributes through $(2\alpha+\beta\zeta_1)$, and the gravitational-sector coefficient $\alpha$ and the matter-coupling coefficient $\beta$ enter the effective combinations. Thus, the observational analysis constrains the net effect of these physical ingredients on the background expansion, while their individual contributions cannot be disentangled using background expansion data alone.

\section{Observational Data Sets}
To constrain the effective parameters of the proposed viscous ghost dark energy model, we perform a joint analysis using three complementary late-time cosmological probes: the Pantheon+SH0ES Type Ia supernova compilation, Cosmic Chronometer (CC) measurements, and Baryon Acoustic Oscillation (BAO) observations from the Dark Energy Spectroscopic Instrument (DESI). These datasets provide independent constraints on the recent expansion history of the Universe over a broad redshift range~\cite{Brout2022,Jimenez2002,Moresco2016,DESI2024}.

\subsection{Type Ia Supernovae}
We use the Pantheon+SH0ES compilation~\cite{Brout2022}, consisting of 1701 spectroscopically confirmed Type Ia supernovae covering the redshift interval $0.001<z<2.3$. Following the official Pantheon+SH0ES likelihood implementation, the analysis is performed using the corrected apparent magnitudes together with the full statistical and systematic covariance matrix released by the Pantheon+ collaboration. The Cepheid-calibrated SH0ES supernovae are treated using their independently measured Cepheid distance moduli, whereas the remaining supernovae are compared with the theoretical apparent magnitudes predicted by the cosmological model.

For the non-calibrator supernovae, the theoretical apparent magnitude is given by
\begin{equation}
	m_{\rm th}(z)= 5\log_{10}\left[(1+z)D_C(z)\right]+25+M,
\end{equation}
where
\begin{equation}
	D_C(z)	=c\int_0^z\frac{dz'}{H(z')}
\end{equation}
is the comoving distance. For the Cepheid-calibrated SH0ES supernovae, the independently measured Cepheid distance moduli are used, and the corresponding theoretical apparent magnitudes are obtained by adding the absolute magnitude parameter $M$. The absolute magnitude $M$ is treated as a nuisance parameter and is constrained simultaneously with the cosmological parameters.

\subsection{Cosmic Chronometers}

The Cosmic Chronometers (CC) method provides direct measurements of
the Hubble parameter through the differential age evolution of
passively evolving galaxies \cite{Jimenez2002,Moresco2016}. The Hubble
parameter is obtained from
\begin{equation}\label{eq29a}
	H(z)=-\frac{1}{1+z}\frac{dz}{dt}.
\end{equation}
In the present analysis, we use a compilation of 36 independent
Cosmic Chronometer measurements of $H(z)$ spanning the redshift range
$0.07\leq z\leq1.965$. The complete set of redshifts, measured Hubble
parameter values, and their corresponding $1\sigma$ uncertainties used
in our analysis is provided in Appendix~\ref{app:CCdata}. The CC
dataset is used without imposing any additional theoretical prior on
the redshift dependence of $H(z)$.

The corresponding chi-square contribution is given by
\begin{equation}\label{eq30}
	\chi^2_{\rm CC}
	=
	\sum_{i=1}^{36}
	\frac{
		\left[
		H_{\rm obs}(z_i)-H_{\rm th}(z_i)
		\right]^2
	}{
		\sigma_{H,i}^2
	}.
\end{equation}
Here, $H_{\rm obs}(z_i)$ and $\sigma_{H,i}$ denote the observed Hubble
parameter and its $1\sigma$ uncertainty at redshift $z_i$, respectively,
while $H_{\rm th}(z_i)$ represents the theoretical prediction of the
model. These measurements provide a direct late-time probe of the
cosmic expansion history and complement the Type Ia supernova and BAO
datasets used in the combined analysis.

\subsection{DESI BAO Data}
We further include the latest BAO observations from the Dark Energy Spectroscopic Instrument (DESI) survey \cite{DESI2024}. BAO measurements provide geometrical constraints through the quantities $\frac{D_M(z)}{r_d}$ and $\frac{D_H(z)}{r_d}$, where
\begin{equation}\label{eq31}
	D_M(z)
	=
	c\int_0^z\frac{dz'}{H(z')}
\end{equation}
denotes the comoving angular diameter distance and
\begin{equation}\label{eq32}
	D_H(z)=\frac{c}{H(z)}
\end{equation}
represents the Hubble distance. The BAO likelihood is constructed using the full DESI covariance matrix. In the present analysis, the sound horizon at the drag epoch, $r_d$, is treated as a free parameter and is constrained simultaneously with the cosmological parameters. This approach avoids assuming the standard $\Lambda$CDM value of $r_d$, which may not be valid in modified gravity theories, and provides a self-consistent analysis of the BAO observations.

The three observational probes are assumed to be statistically independent. Their corresponding likelihoods are combined in the Bayesian analysis described in the next section.

\section{Statistical Analysis}
To constrain the effective parameters of the proposed model, we
confront its theoretical predictions with the available observational
datasets using a Bayesian parameter estimation framework. The methodology employed and the goodness-of-fit analysis are presented in the following subsections.
\subsection{Bayesian Parameter Estimation}

Assuming that the three observational datasets are statistically independent, the total likelihood is constructed as the product of the individual likelihoods. Equivalently, the total chi-square function is given by
\begin{equation}
	\chi^2_{\rm tot}
	=
	\chi^2_{\rm SN}
	+
	\chi^2_{\rm CC}
	+
	\chi^2_{\rm DESI}.
\end{equation}

The posterior distributions of the model parameters are explored using the affine-invariant Markov Chain Monte Carlo (MCMC) ensemble sampler implemented in the \texttt{emcee} package~\cite{ForemanMackey2013}. Since the background expansion history depends only on the effective parameter combinations $(m,n)$, the Bayesian parameter estimation is performed directly in the effective parameter space rather than in terms of the original theoretical parameters. Accordingly, the parameter vector sampled in the MCMC analysis is
\begin{equation}\label{eq34}
	\Theta=
	\left\{
	m,\,
	n,\,
	H_0,\,
	r_d,\,
	M
	\right\},
\end{equation}
where $r_d$ denotes the sound horizon at the drag epoch and $M$ is the absolute magnitude nuisance parameter of Type Ia supernovae.

Uniform priors are assigned to all sampled parameters over physically motivated ranges. The MCMC analysis is performed using 120 walkers and $1.2\times10^4$ sampling steps. Convergence of the Markov chains is assessed using the integrated autocorrelation time together with visual inspection of the trace plots. The total chain length exceeds fifty times the maximum autocorrelation time, indicating satisfactory convergence of the posterior samples. The posterior distributions are summarized by their median values together with the 16th and 84th percentiles, corresponding to the $68\%$ credible intervals. The resulting parameter constraints obtained from the combined Pantheon+SH0ES, Cosmic Chronometers, and DESI BAO analysis are presented in Table~\ref{tab:constraints}.
\begin{table}[htbp]
	\centering
	\caption{Marginalized $68\%$ credible intervals for the effective cosmological parameters obtained from the combined observational analysis. The quoted values correspond to the posterior median together with the 16th and 84th percentiles.}
	\label{tab:constraints}
	\begin{tabular}{lc}
		\hline\hline
		Parameter & $68\%$ credible interval \\
		\hline
		
		$m$
		& $0.706^{+0.020}_{-0.021}$ \\
		
		$n$
		& $1.861^{+0.053}_{-0.052}$ \\
		
		$H_0~(\mathrm{km\,s^{-1}\,Mpc^{-1}})$
		& $71.663^{+0.831}_{-0.822}$ \\
		
		$r_d~(\mathrm{Mpc})$
		& $139.91^{+1.69}_{-1.67}$ \\
		
		$M$
		& $-19.296^{+0.024}_{-0.024}$ \\
		
		\hline\hline
	\end{tabular}
\end{table}
\begin{figure}[htbp]
	\centering
	\includegraphics[width=0.98\linewidth]{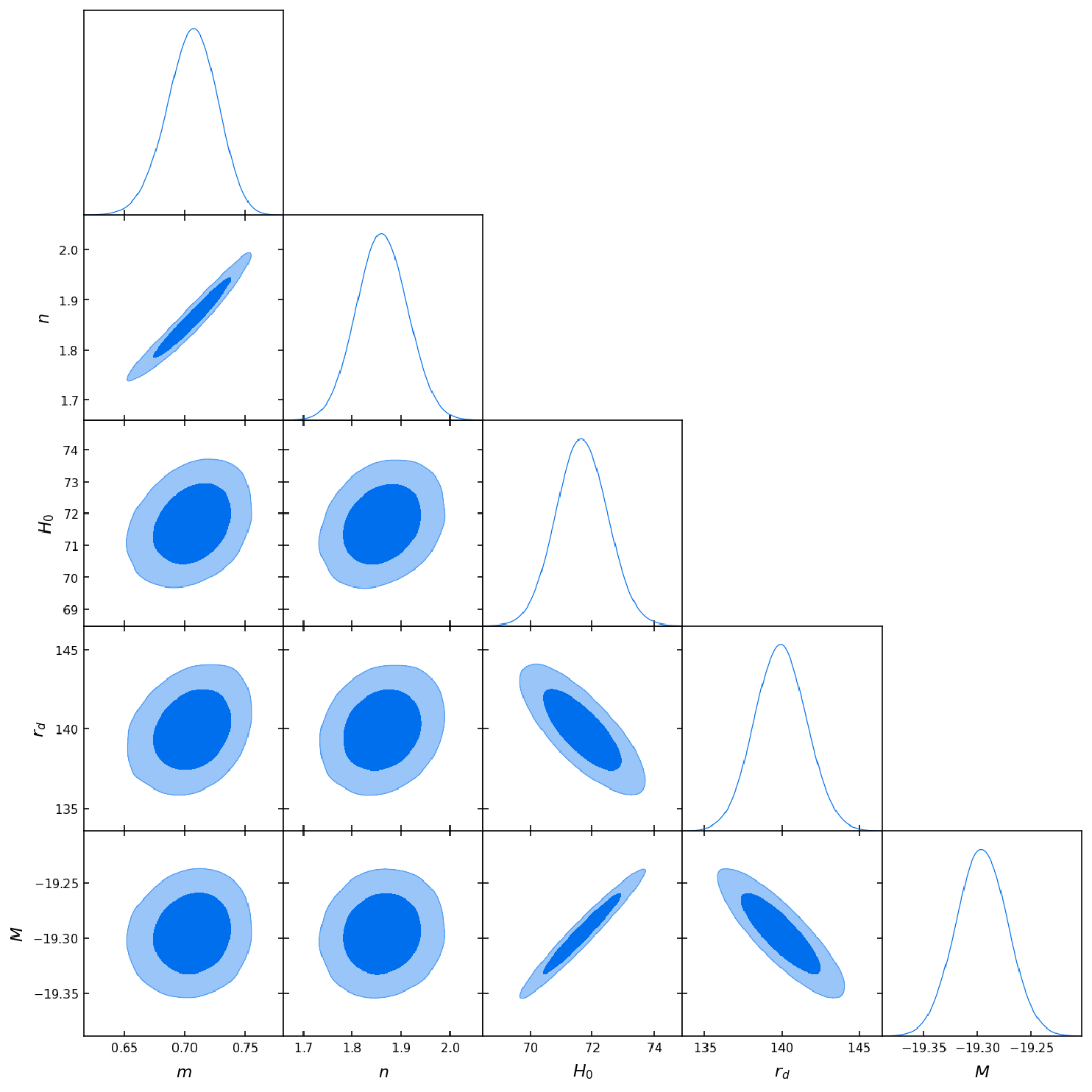}
	\caption{
		One-- and two--dimensional marginalized posterior distributions of the effective cosmological parameters obtained from the combined observational analysis.
	}
	\label{fig:corner}
\end{figure}

Figure~\ref{fig:corner} presents the one-- and two--dimensional marginalized posterior distributions of the sampled parameters obtained from the combined observational analysis. The correspondence between the effective parameters $(m,n)$ and the original theoretical parameters is given by Eq.~(\ref{eq28}).

The joint observational analysis constrains the effective parameters to $m=0.706^{+0.020}_{-0.021}$ and
$n=1.861^{+0.053}_{-0.052}$, indicating that the late-time expansion history is tightly constrained by the combined datasets. The present Hubble parameter is obtained as
$H_0=71.663^{+0.831}_{-0.822}\,\mathrm{km\,s^{-1}\,Mpc^{-1}}$,
while the sound horizon at the drag epoch is constrained to
$r_d=139.91^{+1.69}_{-1.67}\,\mathrm{Mpc}$.
The nuisance parameter associated with the absolute magnitude of Type Ia supernovae is constrained to
$M=-19.296^{+0.024}_{-0.024}$,
consistent with previous analyses of the Pantheon+SH0ES sample.

Overall, the combined observational analysis provides tight constraints on the effective cosmological parameters and demonstrates that the proposed viscous ghost dark energy model successfully reproduces the observed late-time expansion history of the Universe. The posterior samples are subsequently propagated to derive the cosmological quantities $q_0$, $z_t$, and $w_{\rm eff,0}$ together with their corresponding $68\%$ credible intervals, which are discussed in the next section.

\subsection{Goodness-of-fit and Model Comparison}

To assess the statistical performance of the proposed model, we compute the minimum chi-square together with the Akaike Information Criterion (AIC)~\cite{Akaike1974} and the Bayesian Information Criterion (BIC)~\cite{Schwarz1978}. The AIC and BIC are defined, respectively, as
\begin{equation}
	{\rm AIC}=\chi^2_{\rm min}+2k,
\end{equation}
and
\begin{equation}
	{\rm BIC}
	=
	\chi^2_{\rm min}
	+
	k\ln N,
\end{equation}
where $k$ denotes the number of free parameters and $N$ is the total number of observational data points. While both criteria penalize models with a larger number of free parameters, the BIC imposes a stronger complexity penalty than the AIC. Consequently, these statistics provide a useful means of comparing competing cosmological models beyond the minimum chi-square alone.

For model comparison, we evaluate the differences in the information criteria with respect to the flat $\Lambda$CDM model,
\begin{equation}
	\Delta{\rm AIC}
	=
	{\rm AIC}_{\rm model}
	-
	{\rm AIC}_{\Lambda{\rm CDM}},
\end{equation}
and
\begin{equation}
	\Delta{\rm BIC}
	=
	{\rm BIC}_{\rm model}
	-
	{\rm BIC}_{\Lambda{\rm CDM}}.
\end{equation}
Following the model selection criteria discussed in Refs.~\cite{Liddle2004,Liddle2007}, models with $\Delta{\rm AIC}<2$ are considered to have substantial observational support, whereas $4<\Delta{\rm AIC}<7$ indicates considerably weaker support. Similarly, $2<\Delta{\rm BIC}<6$ is generally interpreted as positive evidence against the model with the larger BIC, while $\Delta{\rm BIC}>6$ corresponds to strong evidence against it.

\begin{table}[htbp]
	\centering
	\caption{Comparison of the goodness-of-fit statistics between the present viscous ghost dark energy model and the flat $\Lambda$CDM model using the combined Pantheon+SH0ES, Cosmic Chronometers, and DESI BAO datasets.}
	\label{tab:modelcomparison}
	\begin{tabular}{lcc}
		\hline\hline
		Statistic & Present model & Flat $\Lambda$CDM \\
		\hline
		$\chi^2_{\min}$ & 1487.81 & 1490.95 \\
		$\chi^2_{\nu}$ & 0.875 & 0.877 \\
		AIC & 1497.81 & 1498.95 \\
		BIC & 1525.02 & 1520.71 \\
		\hline\hline
	\end{tabular}
\end{table}

The goodness-of-fit statistics are summarized in Table~\ref{tab:modelcomparison}.
For the proposed model, we obtain a minimum chi-square of
$\chi^2_{\rm min}=1487.81$ and a reduced chi-square of
$\chi^2_{\nu}=0.875$. The latter is close to unity, indicating an
acceptable fit to the combined Pantheon+SH0ES, Cosmic Chronometers, and
DESI BAO datasets. The corresponding statistics for the flat
$\Lambda$CDM model are also reported in the table for comparison.

From Table~\ref{tab:modelcomparison}, we obtain $\Delta{\rm AIC}=-1.14$ and $\Delta{\rm BIC}=4.31$ relative to the flat $\Lambda$CDM model. The small negative value of $\Delta{\rm AIC}$ indicates that the proposed model is slightly favored over the standard cosmological model. In contrast, the value of $\Delta{\rm BIC}$ suggests positive, although not strong, evidence in favor of the simpler $\Lambda$CDM model owing to its smaller number of free parameters. Therefore, while the present model provides a modest improvement in the fit to the late-time observations, this improvement is not sufficient to overcome the stronger complexity penalty imposed by the BIC. Overall, the proposed viscous ghost dark energy model remains statistically competitive with the standard $\Lambda$CDM cosmology while offering a physically motivated extension through bulk viscosity and the non-minimal matter--geometry coupling in $f(Q,\mathcal{L}_m)$ gravity.

\section{Cosmological Evolution and Thermodynamic Viability}

Having constrained the effective model parameters using the observational data, we now investigate the cosmological evolution and physical viability of the model. We examine the evolution of the deceleration parameter, the effective equation-of-state parameter, the matter density parameter, and the Hubble parameter to characterize the expansion history and the transition from decelerated to accelerated expansion. We further analyze the generalized second law of thermodynamics to assess the thermodynamic viability of the proposed viscous ghost dark energy model in $f(Q,L_m)$ gravity.
\subsection{Deceleration Parameter}
The deceleration parameter (DP) provides a direct measure of the expansion dynamics of the Universe and is particularly useful for characterizing the transition from an early decelerated phase to the late-time accelerated expansion. We therefore examine its evolution in the present model. It is defined as
\begin{equation}
	q=-1-\frac{\dot H}{H^2}.
\end{equation}
Using the relation
\begin{equation*}
\dot H=-(1+z)H\frac{dH}{dz},
\end{equation*}
the above expression becomes
\begin{equation}
	q=-1+\frac{(1+z)}{H}\frac{dH}{dz}.
\end{equation}
Substituting the effective Hubble parameter given by Eq.~(\ref{eq29}), we obtain
\begin{equation}
	q(z)
	=
	-1+
	\frac{
		n(1-m)(1+z)^n
	}{
		m+(1-m)(1+z)^n
	}.
\end{equation}
\begin{figure}
	\centering
	\includegraphics[width=0.5\linewidth]{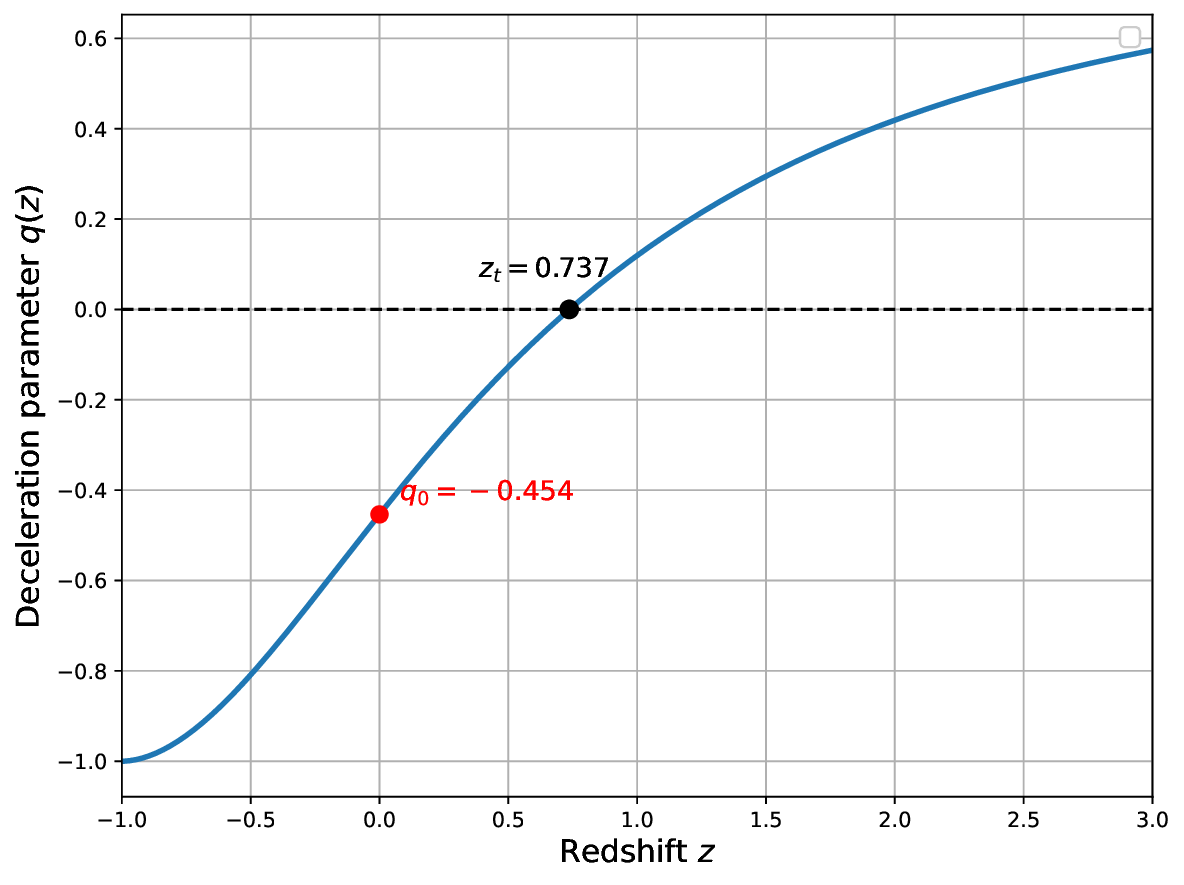}
	\caption{Evolution of the deceleration parameter $q(z)$ as a function of redshift. The black point marks the transition redshift $z_t$ from decelerated to accelerated expansion, while the red point indicates the present-day value $q_0$.}
	\label{fig:q}
\end{figure}
Figure~\ref{fig:q} shows that the Universe evolves from an early decelerating phase ($q>0$) to the present accelerating phase ($q<0$). Propagation of the posterior samples yields $q_0=-0.454^{+0.024}_{-0.024}$, while the transition from deceleration to acceleration occurs at $z_t=0.737^{+0.023}_{-0.024}$, both quoted at the $68\%$ credible level. These values are consistent with recent observational determinations and indicate that the proposed model naturally reproduces the observed transition from a decelerating Universe to the present accelerated phase.

\subsection{Effective Equation of State}

The effective equation-of-state (EoS) parameter provides an important characterization of the dynamical nature of the cosmic expansion and allows us to examine whether the evolution of the Universe approaches, crosses, or departs from the cosmological-constant regime. In the present model, it is therefore useful to investigate the redshift evolution of the effective EoS parameter and its behavior at the present epoch. It is defined as
\begin{equation}
	w_{\rm eff}=\frac{p_{\rm eff}}{\rho_{\rm eff}}.
\end{equation}
\begin{equation}
	\omega_{\rm eff}
	=
	-1+
	\frac{2(1+z)}{3H}
	\frac{dH}{dz}.
\end{equation}
Using Eq.~(\ref{eq29}), it can be written as
\begin{equation}
	\omega_{\rm eff}(z)
	=
	-1+
	\frac{
		\frac{2n}{3}(1-m)(1+z)^n
	}{
		m+(1-m)(1+z)^n
	}.
\end{equation}
\begin{figure}
	\centering
	\includegraphics[width=0.5\linewidth]{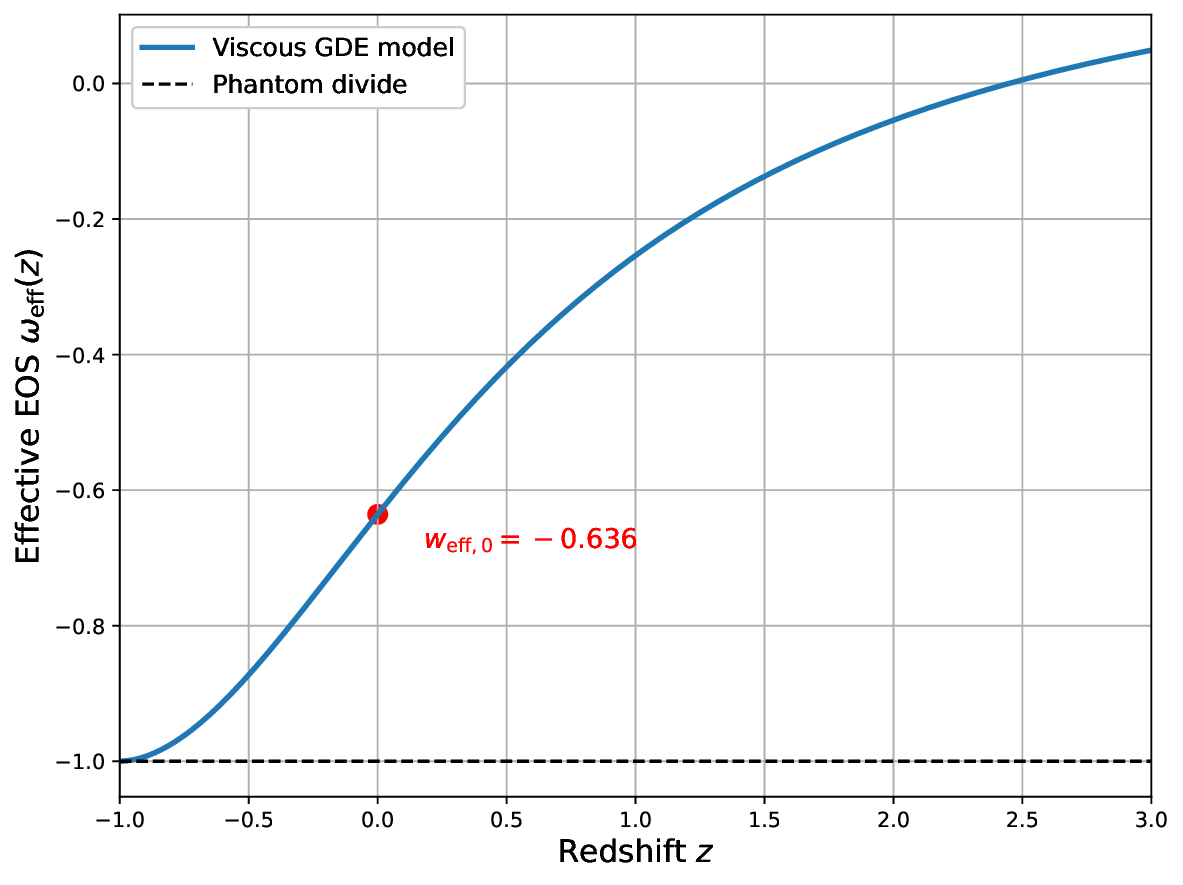}
	\caption{Evolution of the effective EoS parameter $\omega_{\rm eff}(z)$ with redshift. The red point denotes the present-day value $\omega_{\rm eff,0}$.}
	\label{fig:w}
\end{figure}
Fig.~\ref{fig:w} shows that the effective EoS parameter evolves smoothly from values close to the matter-dominated regime at higher redshifts toward negative values at late times. During the late-time future evolution, effective EoS parameter approaches the $\Lambda$CDM value. Propagating the MCMC posterior samples gives the present-day effective equation-of-state parameter as $\omega_{\rm eff,0} = -0.636^{+0.016}_{-0.016}$, at the $68\%$ credible level. Since $\omega_{\rm eff,0}< -1/3$, the model naturally predicts an accelerating Universe at the present epoch.

\subsection{Matter Density Evolution}
The evolution of the matter density provides an important test of the cosmological behavior of the model and allows us to examine how the geometry--matter coupling affects the matter sector. In particular, the
present matter density parameter and its redshift evolution are useful for assessing whether the model yields a physically reasonable matter content over the cosmological history. We therefore investigate the
evolution of the matter density parameter $\Omega_m(z)$ in the present
model.

The total energy density follows directly from the modified Friedmann
equation~(\ref{eq19}) as
\begin{equation}
	\rho=-\frac{6\alpha}{\beta}H^2.
\end{equation}
Since the total energy density is composed of pressureless matter and
ghost dark energy, $\rho=\rho_m+\rho_d$, with $\rho_d=\gamma H$, the
matter density is obtained as
\begin{equation}
	\rho_m=-\frac{6\alpha}{\beta}H^2-\gamma H.
\end{equation}
Consequently, the matter density parameter becomes
\begin{equation}
	\Omega_m(z)
	=
	\frac{\rho_m}{\rho}
	=
	1+\frac{\beta\gamma}{6\alpha H(z)}.
	\label{eq50}
\end{equation}

Since the constrained value of $n$ is positive, Eq.~(\ref{eq29}) implies
that $H(z)\rightarrow\infty$ as $z\rightarrow\infty$. It therefore
follows that
\begin{equation}
	\lim_{z\rightarrow\infty}\Omega_m(z)=1,
\end{equation}
indicating that the model approaches a matter-dominated regime at high
redshift.

Moreover, the positivity of the total energy density requires
$\alpha/\beta<0$ from Eq.~(\ref{eq19}). With $\gamma>0$ and $H>0$, the
ghost dark energy density $\rho_d=\gamma H$ is also positive. For the
constrained values $m<1$ and $n>0$, the Hubble parameter is an increasing
function of redshift, $\frac{dH}{dz}>0$. Consequently, using $\dot H=-(1+z)H\,dH/dz$, the Hubble parameter decreases with cosmic time. Since $\beta\gamma/\alpha<0$, Eq.~(\ref{eq50}) then shows that $\Omega_m$ decreases toward the present epoch as the universe evolves from high redshift to late times. This behavior is consistent with the expected transition from a matter-dominated early universe to a DE dominated late-time universe.

At the present epoch, the matter density parameter is given by
\begin{equation}
	\Omega_{m0}
	=
	1+
	\frac{\beta\gamma}{6\alpha H_0}.
\end{equation}
Since the positivity of the total energy density requires
$\alpha/\beta<0$, while $\gamma>0$ and $H_0>0$, we have
\begin{equation}
	\frac{\beta\gamma}{6\alpha H_0}<0.
\end{equation}
Consequently,
\begin{equation}
	\Omega_{m0}<1.
\end{equation}
Furthermore, requiring a positive matter density,
$\rho_{m0}>0$, gives $\Omega_{m0}>0$. Thus, for a physically admissible
matter sector,
\begin{equation}
	0<\Omega_{m0}<1,
\end{equation}
consistent with a universe containing both matter and a positive DE component at the present epoch.

However, $\Omega_{m0}$ depends explicitly on the combination $\beta\gamma/\alpha$. Since the late-time background expansion constrains only the effective parameter combinations $(m,n)$, this combination cannot be uniquely reconstructed from the present background data alone. Consequently, a unique determination of $\Omega_{m0}$ requires additional observables that probe independent aspects of the cosmological dynamics, such as measurements of the growth of cosmic structure, weak gravitational lensing, or cosmic microwave background observations.

\subsection{Hubble Parameter}

\begin{figure}
	\centering
	\includegraphics[width=0.5\linewidth]{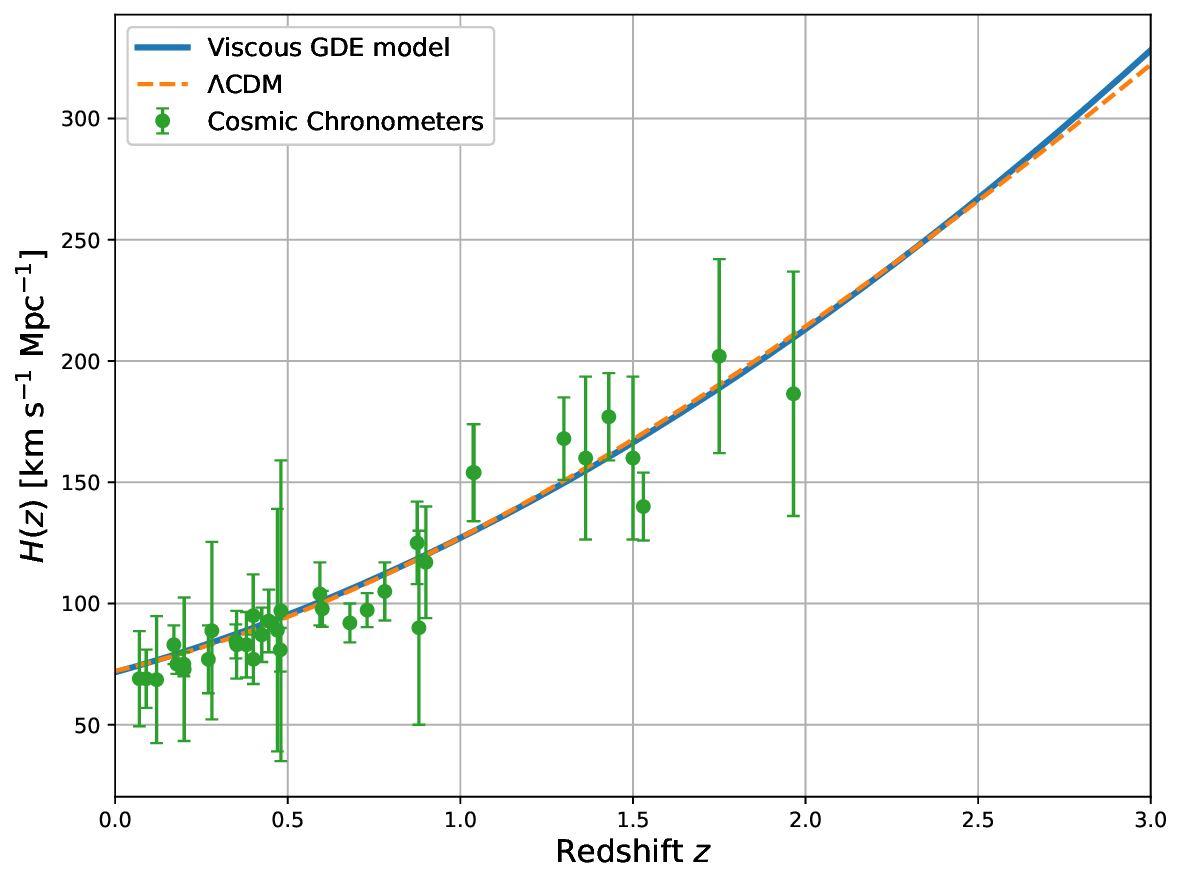}
	\caption{Evolution of the Hubble parameter $H(z)$ for the present viscous ghost dark energy model together with Cosmic Chronometer measurements. The dashed curve corresponds to the best-fit flat $\Lambda$CDM model.}
	\label{fig:H}
\end{figure}

Fig.~\ref{fig:H} compares the theoretical evolution of the Hubble parameter with the Cosmic Chronometer measurements and the best-fit flat $\Lambda$CDM model. The predicted expansion history agrees well with the observational data over the entire redshift range considered. Moreover, only modest deviations from the $\Lambda$CDM prediction are observed at late times, indicating that the present viscous ghost dark energy model provides a viable alternative description of the cosmic expansion history while remaining consistent with current observations.

\subsection{Thermodynamic Analysis}

In addition to its successful description of the background cosmic evolution, the thermodynamic behavior of the proposed model provides an important test of its physical viability. The presence of bulk viscosity
introduces dissipative effects, while the matter--geometry coupling may modify the entropy evolution of the cosmic fluid and the apparent
horizon. We therefore examine the generalized second law of thermodynamics (GSL) to determine whether the total entropy remains non-decreasing during the cosmic evolution. We consider the entropy associated with the apparent horizon together
with the entropy of the effective viscous cosmic fluid enclosed within it. The total entropy is defined as
\begin{equation}
	S_{\rm tot}=S_h+S_m,
	\label{eq:Stot}
\end{equation}
where $S_h$ and $S_m$ denote the horizon and fluid entropies,
respectively. The GSL requires
\begin{equation}
	\dot S_{\rm tot}=\dot S_h+\dot S_m\geq0.
	\label{eq:GSL}
\end{equation}

Before applying the Gibbs relation, we first establish the background
continuity equation for the effective viscous fluid. From
Eqs.~\eqref{eq19} and \eqref{eq20}, we have
\begin{equation}
	\rho=-\frac{6\alpha}{\beta}H^2,
	\label{eq:rho_thermo}
\end{equation}
and
\begin{equation}
	\bar p
	=
	\frac{2\alpha}{\beta}
	\left(2\dot H+3H^2\right).
	\label{eq:pbar_thermo}
\end{equation}
Taking the time derivative of Eq.~\eqref{eq:rho_thermo} gives
\begin{equation}
	\dot{\rho}
	=-\frac{12\alpha}{\beta}H\dot H.
	\label{eq:rhodot_thermo}
\end{equation}
Furthermore,
\begin{equation}
	\rho+\bar p
	=
	-\frac{6\alpha}{\beta}H^2
	+
	\frac{2\alpha}{\beta}
	\left(2\dot H+3H^2\right)
	=
	\frac{4\alpha}{\beta}\dot H.
	\label{eq:rho_pbar}
\end{equation}
It follows from Eqs.~(\ref{eq:rhodot_thermo}) and (\ref{eq:rho_pbar})
\begin{equation}
	3H(\rho+\bar p)
	=
	\frac{12\alpha}{\beta}H\dot H
	=
	-\dot{\rho}.
\end{equation}
Consequently,
\begin{equation}
	\dot{\rho}+3H(\rho+\bar p)=0.
	\label{eq:continuity_thermo}
\end{equation}
Thus, the effective viscous cosmic fluid obeys the standard continuity equation for the homogeneous and isotropic background considered here. This result is important for the thermodynamic analysis. Although energy-momentum conservation is not guaranteed in a general $f(Q,L_m)$ theory, the particular linear model adopted here admits a vanishing temporal source term at the FLRW background level. Thus, the standard continuity equation holds for the effective viscous cosmic fluid, and the Gibbs relation can be consistently applied. 

We now consider the apparent horizon of the spatially flat FLRW universe, whose radius is $r_A=\frac{1}{H}$. Its area is therefore $A=4\pi r_A^2$. Using the Bekenstein--Hawking entropy relation, the horizon entropy is given by
\begin{equation}
	S_h=\frac{A}{4G}
	=\frac{\pi r_A^2}{G}
	=\frac{\pi}{GH^2}.
	\label{eq:Sh}
\end{equation}
Taking the time derivative gives
\begin{equation}
	\dot S_h
	=
	-\frac{2\pi}{G}\frac{\dot H}{H^3}.
	\label{eq:S_h_dot}
\end{equation}
Next, we evaluate the entropy of the effective viscous cosmic fluid
enclosed within the apparent horizon. The energy and volume inside the
horizon are
\begin{equation}
	E=\rho V,
	\qquad
	V=\frac{4\pi}{3}r_A^3.
	\label{eq:E_V}
\end{equation}
For the fluid entropy, we employ the Gibbs relation
\begin{equation}
	T\,dS_m=dE+\bar p\,dV.
	\label{eq:Gibbs}
\end{equation}
Here, the temperature associated with the apparent horizon is taken as
\begin{equation}
	T=\frac{1}{2\pi r_A}
	=\frac{H}{2\pi}.
	\label{eq:T}
\end{equation}
Taking the time derivative of Eq.~\eqref{eq:Gibbs} gives
\begin{equation}
	T\dot S_m
	=
	V\dot\rho+(\rho+\bar p)\dot V.
	\label{eq:Gibbs_dot}
\end{equation}
Using Eq.~\eqref{eq:continuity_thermo} and $\dot V=4\pi r_A^2\dot r_A$, we obtain
\begin{equation}
	\dot S_m
	=
	\frac{4\pi r_A^2}{T}
	(\rho+\bar p)
	\left(\dot r_A-Hr_A\right).
	\label{eq:Sm_general}
\end{equation}
Since
\begin{equation}
	r_A=\frac{1}{H},
	\qquad
	\dot r_A=-\frac{\dot H}{H^2},
	\qquad
	T=\frac{H}{2\pi},
\end{equation}
we obtain
\begin{equation}
	\dot S_m
	=
	-\frac{8\pi^2}{H^3}
	(\rho+\bar p)
	\left(1+\frac{\dot H}{H^2}\right).
	\label{eq:Sm_H}
\end{equation}
Using Eq.~\eqref{eq:rho_pbar}, this becomes
\begin{equation}
	\dot S_m
	=
	-\frac{32\pi^2\alpha}{\beta H^3}
	\dot H
	\left(1+\frac{\dot H}{H^2}\right).
	\label{eq:Sm_final}
\end{equation}
Combining the horizon and fluid contributions, the total entropy production rate is 
\begin{equation}
	\dot S_{\rm tot}
	=
	-\frac{2\pi}{G}\frac{\dot H}{H^3}\left[1
		+\frac{16\pi\alpha G}{\beta}
\left(1+\frac{\dot H}{H^2}\right)\right]
	\label{eq:Stot_q}
\end{equation}
We have the relation
\begin{equation}
\dot H=\frac{dH}{dt}= - (1+z) H \frac{dH}{dz}
\end{equation}
Using Eq.~(\ref{eq29}), we get 

\begin{equation}
	\frac{dH}{dz}
	=
	nH_0(1-m)(1+z)^{n-1},
	\label{eq:dHdz_thermo}
\end{equation}
Consequently,
\begin{equation}
	\dot H
	=
	-nH H_0(1-m)(1+z)^n.
	\label{eq:Hdot_thermo}
\end{equation}
Substituting the values of $H$ and $\dot H$ allows the total entropy production rate to be evaluated directly as a function of redshift. 

It is evident from the expression for $\dot{S}_{\mathrm{tot}}$ that the theoretical parameters $\alpha$ and $\beta$ enter the entropy production
rate through the combination $\alpha/\beta$, while their effects on the background evolution are encoded in the effective parameters $m$ and $n$. Since $\alpha$ and $\beta$ cannot be independently constrained by
the background observational data, the thermodynamic evolution cannot be uniquely reconstructed from the observational constraints alone. Therefore, to illustrate the thermodynamic behavior of the model, we consider a representative parameter combination within the physically
admissible range compatible with the constrained values of $m$ and $n$. \\
\begin{figure}
	\centering
	\includegraphics[width=0.5\linewidth]{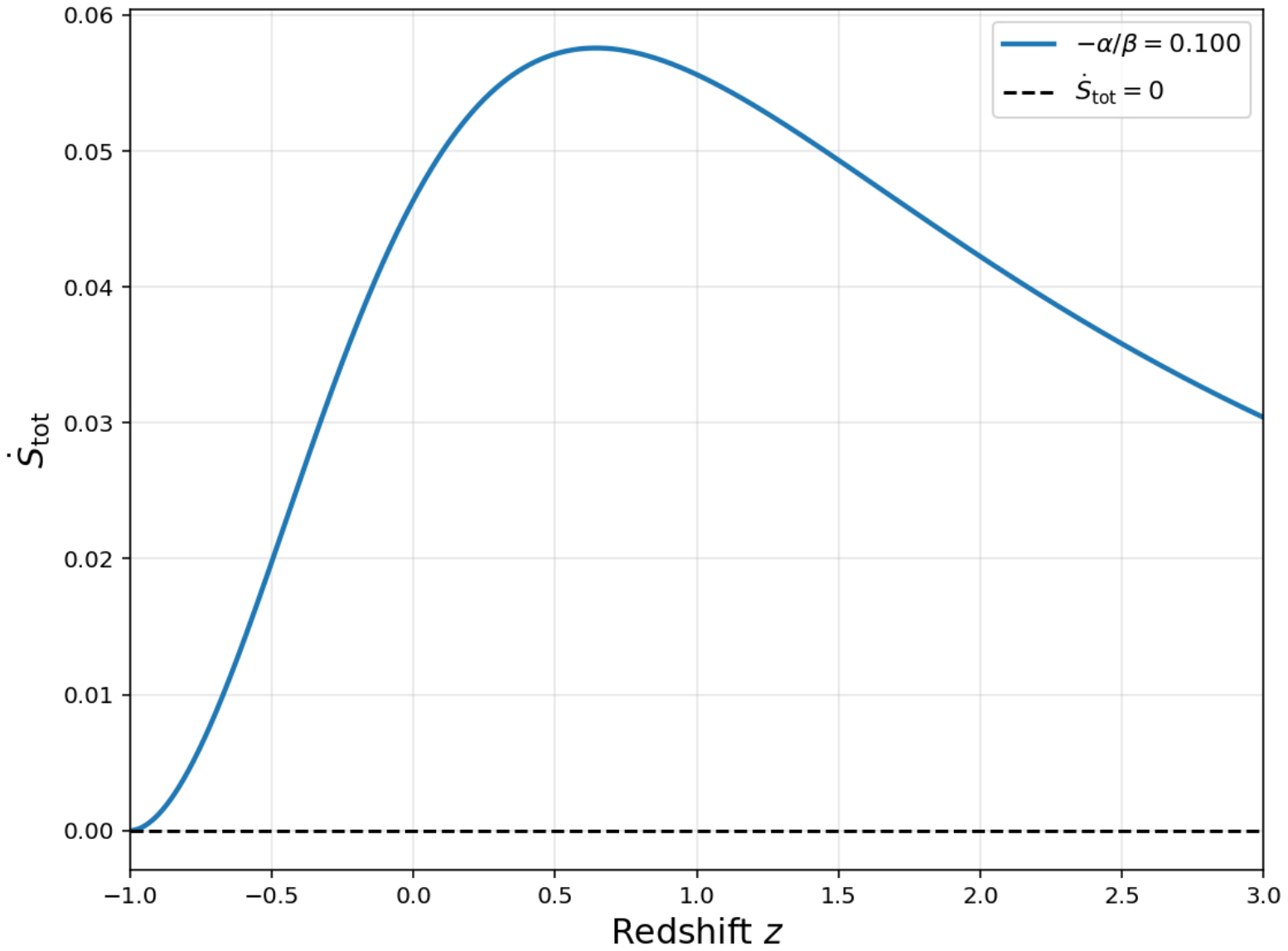}
	\caption{Evolution of the total entropy production rate $\dot{S}_{\mathrm{tot}}$
			as a function of redshift for the representative physically admissible
			choice $\alpha/\beta=-0.1$. The positive value of $\dot{S}_{\mathrm{tot}}$
			throughout the considered redshift range indicates that the generalized
			second law is satisfied for this parameter choice.}
	\label{fig:Stot}
\end{figure}
Figure~\ref{fig:Stot} shows the evolution of the total entropy production rate over the redshift range $-1\leq z\leq3$ for the representative physically admissible choice $\alpha/\beta=-0.1$. The entropy production rate remains positive throughout the considered cosmic
evolution, indicating that the GSL is satisfied for this parameter choice. It reaches a maximum at an intermediate redshift close to the deceleration--acceleration transition, while remaining positive throughout both the early and late stages of cosmic evolution.
\section{Conclusion}
In this work, we investigated the late-time evolution of the Universe by considering a viscous GDE model within the framework of $f(Q,L_m)$ gravity. The presence of bulk viscosity introduces dissipative effects into the cosmic fluid, while the non-minimal coupling between matter and geometry modifies the gravitational dynamics and provides an alternative mechanism for explaining the accelerated expansion of the Universe. Modified cosmological field equations for a spatially flat FLRW Universe have been derived for a linear form of the gravitational Lagrangian and bulk viscosity coefficient $\zeta=\zeta_0+\zeta_1H$. The modified cosmological field equations have been solved to obtain an analytical expression for the Hubble parameter, which was then used
to constrain the effective model parameters using the combined Pantheon+SH0ES Type Ia supernovae, Cosmic Chronometer, and DESI BAO
datasets through an MCMC analysis.

The observational analysis has been formulated in terms of the effective
parameter combinations $(m,n)$, since the background expansion history
depends on the underlying theoretical parameters through these
combinations. The cosmological evolution obtained from the constrained
background expansion exhibits a transition from an early decelerated phase to the present accelerated epoch. The present-day deceleration
parameter is found to be $q_0=-0.454^{+0.024}_{-0.024}$, while the
deceleration--acceleration transition occurs at
$z_t=0.737^{+0.023}_{-0.024}$. The effective EoS parameter also evolves toward negative values at late times, with its present-day
value given by $\omega_{\rm eff}(0)\approx-0.636^{+0.016}_{-0.016}$, consistent with the accelerated expansion. The predicted Hubble parameter follows the Cosmic Chronometer measurements well and remains close to the
best-fit flat $\Lambda$CDM expansion history at late times.

The statistical comparison further shows that the proposed model provides a slightly lower minimum chi-square than the flat $\Lambda$CDM
model, while the AIC indicates comparable statistical support. The BIC, which imposes a stronger penalty on additional parameters, mildly favors the simpler $\Lambda$CDM model. Thus, the present model remains statistically competitive with the standard cosmological scenario while
providing a physically motivated extension through bulk viscosity and matter--geometry coupling.

We also examined the thermodynamic viability of the model through the GSL. For the representative physically admissible choice $\alpha/\beta=-0.1$, the total entropy production rate remains positive throughout the considered cosmic evolution, indicating that the GSL is satisfied for this parameter choice. The entropy production reaches its maximum around the transition epoch, indicating that dissipative processes become most significant during the onset of accelerated expansion.

Overall, the results indicate that viscous GDE in $f(Q,L_m)$ gravity can provide a viable description of the observed late-time accelerated expansion while remaining compatible with the
current background observational data. The combined effects of bulk viscosity and matter--geometry coupling offer a physically interesting
extension of the standard cosmological framework. A natural direction for future work is to investigate the perturbation dynamics and growth
of cosmic structure, which may provide additional constraints on the underlying theoretical parameters and help break the degeneracies that remain at the background level. Further constraints from weak lensing, CMB observations, and other complementary probes could also be useful for testing the model beyond the homogeneous background evolution.

\appendix
\section{Cosmic Chronometer Data}
\label{app:CCdata}

For completeness, Table~\ref{tab:CCdata} lists the complete set of
36 Cosmic Chronometer measurements used in the observational analysis.
The three columns correspond to the redshift $z$, the measured Hubble
parameter $H_{\rm obs}(z)$, and its quoted $1\sigma$ uncertainty
$\sigma_H$, respectively. The values are given in
$\mathrm{km\,s^{-1}\,Mpc^{-1}}$.

\begin{table*}[htbp]
	\centering
	\caption{The 36 Cosmic Chronometer measurements used in the analysis.
		Here, $z$ denotes the redshift, $H_{\rm obs}(z)$ is the observed
		Hubble parameter, and $\sigma_H$ is its quoted $1\sigma$ uncertainty.
		The values of $H_{\rm obs}(z)$ and $\sigma_H$ are given in
		$\mathrm{km\,s^{-1}\,Mpc^{-1}}$.}
	\label{tab:CCdata}
	
	\begin{tabular*}{\textwidth}{@{\extracolsep{\fill}}ccc@{\qquad}ccc@{}}
		\hline\hline
		$z$ & $H_{\rm obs}(z)$ & $\sigma_H$
		&
		$z$ & $H_{\rm obs}(z)$ & $\sigma_H$ \\
		\hline
		0.0700  & 69.0  & 19.6
		& 0.5930  & 104.0 & 13.0 \\
		
		0.0900  & 69.0  & 12.0
		& 0.6000  & 97.8  & 7.4 \\
		
		0.1200  & 68.6  & 26.2
		& 0.6800  & 92.0  & 8.0 \\
		
		0.1700  & 83.0  & 8.0
		& 0.7300  & 97.3  & 7.0 \\
		
		0.1790  & 75.0  & 4.0
		& 0.7810  & 105.0 & 12.0 \\
		
		0.1990  & 75.0  & 5.0
		& 0.8750  & 125.0 & 17.0 \\
		
		0.2000  & 72.9  & 29.6
		& 0.8800  & 90.0  & 40.0 \\
		
		0.2700  & 77.0  & 14.0
		& 0.9000  & 117.0 & 23.0 \\
		
		0.2800  & 88.8  & 36.6
		& 1.0370  & 154.0 & 20.0 \\
		
		0.3500  & 84.4  & 7.0
		& 1.0400  & 154.0 & 20.0 \\
		
		0.3520  & 83.0  & 14.0
		& 1.3000  & 168.0 & 17.0 \\
		
		0.3802  & 83.0  & 13.5
		& 1.3630  & 160.0 & 33.6 \\
		
		0.4000  & 95.0  & 17.0
		& 1.4300  & 177.0 & 18.0 \\
		
		0.4004  & 77.0  & 10.2
		& 1.5000  & 160.0 & 33.6 \\
		
		0.4247  & 87.1  & 11.2
		& 1.5300  & 140.0 & 14.0 \\
		
		0.44497 & 92.8  & 12.9
		& 1.7500  & 202.0 & 40.0 \\
		
		0.4700  & 89.0  & 50.0
		& 1.9650  & 186.5 & 50.4 \\
		
		0.4783  & 80.9  & 9.0
		& & & \\
		
		0.4800  & 97.0  & 62.0
		& & & \\
		
		\hline\hline
	\end{tabular*}
\end{table*}
\section*{Acknowledgements}

The authors are grateful to the anonymous reviewer for the constructive comments and suggestions, which have helped improve the manuscript in its present form. The authors sincerely acknowledge Prof. C. P. Singh, Delhi Technological University, India, for his valuable suggestions, constructive discussions, and continuous encouragement. Pankaj Kumar gratefully acknowledges the Inter-University Centre for Astronomy and Astrophysics (IUCAA), Pune, India, for providing research facilities and support under the Visiting Associateship Programme.


\begin{thebibliography}{}
\bibitem{Riess1998}
A.~G.~Riess \textit{et al.},
``Observational Evidence from Supernovae for an Accelerating Universe and a Cosmological Constant,''
Astron.\ J.\ \textbf{116}, 1009 (1998).

\bibitem{Perlmutter1999}
S.~Perlmutter \textit{et al.},
``Measurements of Omega and Lambda from 42 High-Redshift Supernovae,''
Astrophys.\ J.\ \textbf{517}, 565 (1999).

\bibitem{Spergel2007}
D.~N.~Spergel \textit{et al.},
``Three-Year Wilkinson Microwave Anisotropy Probe (WMAP) Observations: Implications for Cosmology,''
Astrophys.\ J.\ Suppl.\ Ser.\ \textbf{170}, 377 (2007).

\bibitem{Planck2020}
Planck Collaboration,
``Planck 2018 results. VI. Cosmological parameters,''
Astron.\ Astrophys.\ \textbf{641}, A6 (2020).

\bibitem{Tegmark2004}
M.~Tegmark \textit{et al.},
``Cosmological parameters from SDSS and WMAP,''
Phys.\ Rev.\ D \textbf{69}, 103501 (2004).

\bibitem{Eisenstein2005}
D.~J.~Eisenstein \textit{et al.},
``Detection of the Baryon Acoustic Peak in the Large-Scale Correlation Function of SDSS Luminous Red Galaxies,''
Astrophys.\ J.\ \textbf{633}, 560 (2005).

\bibitem{Weinberg1989}
S.~Weinberg,
``The Cosmological Constant Problem,''
Rev.\ Mod.\ Phys.\ \textbf{61}, 1 (1989).

\bibitem{Copeland2006}
E.~J.~Copeland, M.~Sami and S.~Tsujikawa,
``Dynamics of Dark Energy,''
Int.\ J.\ Mod.\ Phys.\ D \textbf{15}, 1753 (2006).

\bibitem{Sahni2014}
V.~Sahni and A.~Starobinsky,
``The Case for a Positive Cosmological Lambda-Term,''
Int.\ J.\ Mod.\ Phys.\ D \textbf{9}, 373 (2000).

\bibitem{Riess2022}
A.~G.~Riess \textit{et al.},
``A Comprehensive Measurement of the Local Value of the Hubble Constant with 1 km/s/Mpc Uncertainty,''
Astrophys.\ J.\ Lett.\ \textbf{934}, L7 (2022).

\bibitem{DESI2024}
DESI Collaboration,
``DESI 2024 Results: Constraints on Dark Energy from the First Year of Data,''
JCAP \textbf{02}, 021 (2025).

\bibitem{Urban2010}
F.~R.~Urban and A.~R.~Zhitnitsky,
``The QCD nature of dark energy,''
Nucl.\ Phys.\ B \textbf{835}, 135 (2010).

\bibitem{Ohta2011}
N.~Ohta,
``Dark energy and QCD ghost,''
Phys.\ Lett.\ B \textbf{695}, 41 (2011).

\bibitem{Veneziano1979}
G.~Veneziano,
``$U(1)$ without instantons,''
Nucl.\ Phys.\ B \textbf{159}, 213 (1979).

\bibitem{Cai2012}
R.~G.~Cai \textit{et al.},
``Notes on ghost dark energy,''
Phys.\ Rev.\ D \textbf{84}, 123501 (2011).

\bibitem{nojiri2011}
S. Nojiri and S.~D. Odintsov
``Unified cosmic history in modified gravity: from $f(R)$ theory to Lorentz non-invariant models''
Phys. Rept. {\bf505}, 59 (2011).

\bibitem{harko2011}
T. Harko, et al.,
``$f(R,T)$ gravity''
Phys. Rev. D {\bf84}, 024020 (2011).

\bibitem{felice2009}
A. De Felice and S. Tsuijikawa,
``Construction of cosmologically viable $f(G)$ gravity models''
Phys. Lett. B {\bf675}, 1 (2009).

\bibitem{jimenez2018}
J.~B.~Jimenez, L.~Heisenberg and T.~Koivisto,
``Coincident General Relativity,''
Phys.\ Rev.\ D \textbf{98}, 044048 (2018).

\bibitem{xu2019}
Y. Xu, G. Li, T. Harko and S.D. Liang,
``$f(Q,T)$ gravity''
Eur. Phys. J. C {\bf79}, 708 (2019).

\bibitem{Nester1998}
J.~M.~Nester and H.~J.~Yo,
``Symmetric teleparallel general relativity,''
Chinese J.\ Phys.\ \textbf{37}, 113 (1999).

\bibitem{Adak2006}
M.~Adak,
``The symmetric teleparallel gravity,''
Turkish J.\ Phys.\ \textbf{30}, 379 (2006).

\bibitem{Jarv2018}
L.~J\"arv, M.~R\"unkla, M.~Saal and O.~Vilson,
``Nonmetricity formulation of general relativity and its scalar--tensor extension,''
Phys.\ Rev.\ D \textbf{97}, 124025 (2018).

\bibitem{Hazarika2025}
A.~Hazarika, S.~Arora, P.~K.~Sahoo and T.~Harko,
``$f(Q,L_m)$ gravity and its cosmological implications,''
Phys.\ Dark Univ.\ \textbf{50}, 102092 (2025).

\bibitem{ilg1999}
 P. Ilg and H. C. \"{O}ttinger, ``Nonequilibrium relativistic thermodynamics in bulk viscous cosmology,''
 Phys. Rev. D {\bf 61}, 023510 (1999).

\bibitem{wilson2007}
 J. R. Wilson, G. J. Mathews and G. M. Fuller, ``Bulk viscosity, decaying dark matter, and the cosmic acceleration,''
 Phys. Rev. D {\bf 75}, 043521 (2007).
\bibitem{eckart1940} C. Eckart, ``The Thermodynamics of Irreversible Processes. III. Relativistic Theory of the Simple Fluid'', Phys. Rev. {\bf58}, 919 (1940).
\bibitem{israel1979} W. Israel and J. Stewart, ``Transient relativistic thermodynamics and kinetic theory'',
 Ann. Phys. {\bf118}, 341 (1979).
\bibitem{fabris2006} J. C. Fabris, S.V.B. Gonalves,  R. de S. Ribeiro, ``Bulk viscosity driving the acceleration of the Universe'', Gen. Relativ. Gravit. {\bf38}, 495 (2006).
\bibitem{ren2006}  J. Ren and X. -He  Meng,  ``Dark viscous fluid described by a unified equation of state in cosmology'', Phys. Lett. B {\bf633}, 1 (2006).
\bibitem{meng2007} X. H. Meng, J. Ren and M. G. Hu, ``Friedmann cosmology with a generalized equation of state and bulk viscosity'' , Commun. Theor. Phys. {\bf47}, 379 (2007).
\bibitem{singh2007a} C. P. Singh, S. Kumar and A. Pradhan, ``Early Viscous Universe with Variable Gravitational and Cosmological Constants'',  Class. Quantum Grav. {\bf24}, 455 (2007).
\bibitem{singh2007b} C. P. Singh, ``Viscosity Effects on the Early Universe with Time-Varying Constants'', Nuovo Cim. B, {\bf122}, 89  (2007).
\bibitem{avelino2008} A. Avelino and U. Nucamendi, ``Can a matter-dominated model with constant bulk viscosity drive the accelerated expansion of the universe?'', J. Cosmol. Astropart. Phys. {\bf04}, 006 (2009).
\bibitem{avelino2010} A. Avelino and U. Nucamendi, ``Exploring a matter-dominated model with bulk viscosity to drive the accelerated expansion of the Universe'', J. Cosmol. Astropart. Phys. {\bf08}, 009 (2010).
\bibitem{avelino2013} A. Avelino,  et al., ``Bulk viscous matter-dominated Universes: asymptotic properties'',  J. Cosmol. Astropart. Phys. {\bf08}, 012 (2013).
\bibitem{singh2014} C. P. Singh and P. Kumar, ``Friedmann model with viscous cosmology in modified $f(R,T)$ gravity theory'',  Eur. Phys. J. C {\bf74}, 3070  (2014).
\bibitem{singh2018a} C. P. Singh and M. Srivastava, ``Viscous cosmology in new holographic dark energy model and the cosmic acceleration'', Eur. Phys. J. C {\bf78}, 190 (2018).
\bibitem{singh2018b}  C. P. Singh and A. Kumar, ``Ricci dark energy model with bulk viscosity'', Eur. Phys. J. Plus {\bf133}, 312 (2018a).
\bibitem{singh2018c}  C. P. Singh and A. Kumar, ``Viscous Ricci dark energy and generalized second law of
thermodynamics in modified $f(R, T )$ gravity'',  Mod. Phys. Lett. A {\bf33},  1850225, (2018b).
\bibitem{singh2019} C. P. Singh and A. Kumar, ``Observational constraints on holographic dark energy model with
matter creation'',  Astrophys. Space Sci. {\bf364} 94 (2019).
\bibitem{singh2021} A. Kumar and C.P. Singh, ``The generalized second law of thermodynamics in viscous Ricci dark energy model'',  Eur. Phys. J. Plus {\bf136}, 820 (2021).
\bibitem{singh2023} S. Kaur and C.P. Singh, ``Viscous cosmology in holographic dark energy with Granda-Oliveros cut-off'', Commun. Theor. Phys. {\bf75}, 025401, (2023).
\bibitem{singh2024} C. P. Singh and V. Khatri, ``Viscous fluid dynamics with decaying vacuum energy density'', Phys. Rev. D {\bf109}, 023508 (2024).
\bibitem{singh2025} V. Khatri and C. P.  Singh, ``Interacting model of bulk viscous and decaying vacuum energy'', Phys. Lett. B {\bf871}, 139994 (2025).
\bibitem{Myrzakulov2024a}
Y.~Myrzakulov \textit{et al.},
``Late-time cosmology in $f(Q,L_m)$ gravity: Analytical solutions and observational fits,''
Phys.\ Dark Univ.\ \textbf{46}, 101614 (2024).

\bibitem{Myrzakulov2024b}
K.~Myrzakulov \textit{et al.},
``Observational analysis of late-time acceleration in $f(Q,L_m)$ gravity,''
JHEAp \textbf{44}, 164 (2024).

\bibitem{Myrzakulov2025}
Y.~Myrzakulov \textit{et al.},
``Energy conditions in $f(Q,L_m)$ gravity,''
Eur.\ Phys.\ J.\ C \textbf{85}, 376 (2025).

\bibitem{Rana2025}
D.~S.~Rana, S.~S.~Mishra, A.~Bhat and P.~K.~Sahoo,
``Viscous Cosmology in $f(Q,L_m)$ Gravity: Insights from CC, BAO, and GRB Data,''
Universe \textbf{11}, 242 (2025).

%

\bibitem{Hehl1995}
F.~W.~Hehl, J.~D.~McCrea, E.~W.~Mielke and Y.~Ne'eman,
``Metric-affine gauge theory of gravity,''
Phys.\ Rept.\ \textbf{258}, 1 (1995).

\bibitem{landau1987} L. D. Landau and E. M. Lifshitz,  Fluid Mechanics (Butterworth Heinemann, Oxford, (1987).

\bibitem{Singh2008}
C.~P.~Singh,
``Bulk viscous cosmological models,''
Pramana J.\ Phys.\ \textbf{71}, 33 (2008).

\bibitem{Meng2009}
X.~H.~Meng and X.~Dou,
``Viscous cosmological models and dark energy,''
Commun.\ Theor.\ Phys.\ \textbf{52}, 377 (2009).

\bibitem{Brout2022}
D.~Brout \textit{et al.},
``The Pantheon+ Analysis: Cosmological Constraints,''
Astrophys.\ J.\ \textbf{938}, 110 (2022).

\bibitem{Jimenez2002}
R.~Jimenez and A.~Loeb,
``Constraining cosmological parameters based on relative galaxy ages,''
Astrophys.\ J.\ \textbf{573}, 37 (2002).

\bibitem{Moresco2016}
M.~Moresco \textit{et al.},
``A 6\% measurement of the Hubble parameter at $z\sim0.45$,''
JCAP \textbf{05}, 014 (2016).

\bibitem{ForemanMackey2013}
D.~Foreman-Mackey, D.~W.~Hogg, D.~Lang and J.~Goodman,
``emcee: The MCMC Hammer,''
Publ.\ Astron.\ Soc.\ Pac.\ \textbf{125}, 306 (2013).

\bibitem{Akaike1974}
H. Akaike, ``A New Look at the Statistical Model Identification'',
IEEE Transactions on Automatic Control \textbf{19}, 716 (1974).

\bibitem{Schwarz1978}
G. Schwarz,
``Estimating the Dimension of a Model'',
The Annals of Statistics \textbf{6}, 461 (1978)

\bibitem{Liddle2004}
A.R. Liddle,
``How many cosmological parameters?'',
Monthly Notices of the Royal Astronomical Society \textbf{351},
L49 (2004)

\bibitem{Liddle2007}
A.R. Liddle,
``Information criteria for astrophysical model selection'',
Monthly Notices of the Royal Astronomical Society \textbf{377},
L74 (2007)

\end{thebibliography}
\end{document}